\documentclass[prb,twocolumn,showpacs,preprintnumbers,amsmath,amssymb]{revtex4}

\usepackage{graphicx}% Include figure files
\usepackage{subfigure} %To include 2 figures in one column
\usepackage{dcolumn}% Align table columns on decimal point
\usepackage{bm}% bold math

\begin{document}

\draft

\title{Inelastic carrier lifetime in graphene}
\author{E.~H.~Hwang,$^1$ Ben Yu-Kuang Hu,$^{2,1}$ and S.~Das Sarma$^1$}
\address{$^1$Condensed Matter Theory Center, 
Department of Physics, University of Maryland, College Park,
Maryland  20742-4111} 
\address{$^2$Department of Physics,
University of Akron, Akron, OH 44325-4001}
\date{\today}

\begin{abstract}
We consider hot carrier inelastic scattering due to electron--electron
interactions in graphene, as functions of carrier energy and density.
We calculate the imaginary part of the zero-temperature quasiparticle self-energy for doped graphene, utlizing the $G_0W$ and random phases approximations.
Using the full dynamically screened Coulomb interaction, we obtain the inelastic quasiparticle lifetimes and associated mean free paths. 
The linear dispersion of graphene gives lifetime energy dependences that are qualitatively different from those of parabolic-band semiconductors.
We also get good agreement with data from angle-resolved photoemission spectroscopy experiments.

\end{abstract}
\maketitle

\section{introduction}

Graphene, a single layer of carbon atoms covalently bonded together in a honeycomb structure (as in a monolayer of graphite), was previouly thought
to be unstable and hence non-existent in a free state. Recently, however, Novoselov {\em et al.} reported \cite{Geim} that they had succeeded in
fabricating single graphene sheets. Subsequently, several experimental groups have reported interesting transport and spectroscopic 
measurements \cite{Measurements,Bostwick,Lanzara}, which has led to experimental and theoretical interest in this field that is rapidly burgeoning.

The overlap of the $\pi_z$ orbitals between neighboring carbon atoms in the graphene plane 
is accurately described by a tight-binding Hamiltonian.  Around the $K$ and $K'$ points (often called Dirac points)
which are at the corners of the hexagonal Brillouin zone, the kinetic energy term of the Hamiltonian is well-approximated by a two-dimensional (2D)
Dirac equation for massless particles, $\hat{H}_0 = -v_0 (\sigma_x \hat{k}_x + \sigma_y \hat{k}_y)$,
where $\sigma_{x}$ and $\sigma_y$ are $2\times 2$ Pauli spinors and ${\bm k}$ is the momentum relative to the Dirac points ($\hbar = 1$ throughout this paper). 
The two components of the spinors correspond to occupancy of the two sublattices of the honeycomb structure in a hexagonal lattice.  
This $\hat{H}_0$ gives a linear energy dispersion relation $\epsilon_{\bm k,s} = s  v_0 |{\bm k}|$, 
where $s = +1$ $(-1)$ for the conduction (valence) band.
The corresponding density of states (DOS) is 
$ D(\epsilon) = g_s g_v |\epsilon|/(2\pi v_0^2)$, where
$g_s=2$, $g_v=2$ are the spin and valley ({\em i.e.}, $K$ and $K'$ points) degeneracies, respectively.
Thus, graphene is a gapless semiconductor. In intrinsic graphene, the Fermi level lies at the Dirac points, but as with
other semiconductors it is possible to shift the Fermi level 
either by doping the sample or applying an external gate voltage, which introduces 2D free carriers (electrons or holes) producing extrinsic graphene with
gate voltage induced tunable carrier density.
The Fermi momentum ($k_F$) and the Fermi energy ($E_F$, relative to the Dirac point energy) 
of graphene are given by $k_F = (4\pi n/g_s g_v)^{1/2}$ and $|E_F| = v_0 k_F$ where
$n$ is the 2D carrier (electron or hole) density.

Interparticle interactions can significantly affect electronic properties, particularly in systems of reduced dimensionality.
Moreover, the linear energy dispersion around the Dirac points gives condensed matter experimentalists a unique
opportunity to study interaction effects on effectively massless particles.
In this paper, we focus on the effect of electron--electron ($e$--$e$) interaction effects on the imaginary part of quasiparticle self-energies, ${\rm Im}[\Sigma]$.
From ${\rm Im}[\Sigma]$, we can extract the quasiparticle lifetime, which gives information that is relevant both to fundamental questions, 
such as whether or not the system is a Fermi liquid, and to possible applications, 
such as the energy dissipation rate of injected carriers in a graphene-based device. 
In particular, an important physical quantity of both fundamental and technological significance is the hot carrier mean free path,
which we calculate as a functions of energy, density and in-plane
dielectric constant. Furthermore, ${\rm Im}[\Sigma]$, being the width
of the quasiparticle spectral function, is
related to measurements in angle resolved photoemission spectroscopy (ARPES).

The rest of the paper is organized as follows. In section II we
discuss the general theory of self-energy of graphene. In section III
we present our calculated quasiparticle damping rate of graphene and
compare with the damping rate of parabolic 2D systems.
In section IV we show the detail reuslts of the self-energy, and
finally we conclude in section V. 

\section{theory}

We evaluate the self-energy $\Sigma$ within the leading-order ring-diagram $G_0W$ approximation, which is appropriate for weak-coupling
systems, given by\cite{Mahan} 
\begin{eqnarray}
\Sigma_s(\bm k,i\omega_n) =
-k_BT\sum_{s'}\sum_{{\bm q},i\nu_n} 
G_{0,s'}(\bm k+\bm q,i\omega_n+i\nu_n) \nonumber \\
\times W(q,i\nu_n)
F_{ss'}({\bm k},{\bm k}+{\bm q}).
\label{eq:1}
\label{sigma}
\end{eqnarray}
Here, $T$ is temperature, $s,s'=\pm 1$ are band indices, 
$G_{0}$ %({\bf k},i\omega_n)=1/(i\omega_n-\xi_{{\bf k}s})$
is the bare Green's function, $\omega_n,\nu_n$ are Matsubara fermion and boson 
frequencies, respectively, 
$W$ is the screened Coulomb interaction, and $F_{ss'}(\bm{k},\bm{k}')
= \frac{1}{2}(1 + ss' \cos\theta_{\bm{kk}'})$, where  
$\theta_{\bm{kk'}}$ is the angle between $\bm{k}$, $\bm{k}'$,  
arises from the overlap of $\mathinner|\!s{\bm k}\rangle$ and
$\mathinner|\!s'{\bm k}'\rangle$. 
The screened interaction $W(q,i\nu_n) =
V_c(q)/\epsilon(q,i\nu_n)$,  where $V_{c}(q)=2\pi e^2/\kappa q$ is
the bare Coulomb potential  
($\kappa = $background dielectric constant), and
$\varepsilon(q,i\nu_n)$ is the 2D dynamical dielectric function. 
In the random phase approximation $\varepsilon(q,i\nu_n) = 1-V_c(q)
\Pi_0(q,i\nu_n),$  
where the irreducible polarizability $\Pi_0$ is approximated by the
bare bubble diagram \cite{Hwang,PhysRevB.34.979}, which gives the
familiar  
Lindhard expression [with a modification to include the form factor
$F_{ss'}(\bm k,\bm k')$].  

The self-energy approximation described by  Eq.~(\ref{eq:1})
should be an excellent approximation for graphene since graphene is inherently a
weak-coupling 
(or ``high-density" in parabolic-band systems) 2D system
\cite{sarma:121406}. 
The measure of the ratio of the potential to the kinetic energy in graphene is $r_s= e^2/(\kappa v_0) \approx 2.2/\kappa$, where $\kappa$ is
the effective dielectric constant of the graphene layer and the media surrounding it.  
(Note that, unlike regular parabolic-band systems, $r_s$ in graphene does not depend on density.)  
Typically in graphene, $r_s < 1$, implying that the system is in the weak-coupling regime.

After the standard procedure of analytical continuation from
$i\omega_n$ to $\omega + i0^+$, the retarded
self-energy can be separated into the exchange and correlation
parts $ \Sigma_{s}^{\rm ret}({\bf k},\omega) = \Sigma_{s}^{\rm ex}({\bf k})
+ \Sigma_{s}^{\rm cor}({\bf
  k},\omega). $\cite{Mahan}
The exchange part is given by
\begin{equation}
\Sigma^{\rm ex}_{s}({\bf k}) = -\sum_{s'{\bf q}}
\ n_F(\xi_{{\bf k}+{\bf q},s'})\, V_c({\bf q})\, F_{ss'}({\bf
k}, {\bf k}+{\bf q}),
%\label{eq:1}
\end{equation}
where $n_F$ is the Fermi function, and $\xi_{\bm k,s} = \epsilon_{\bm k,s} - \mu$ is the electron
energy relative to the chemical potential $\mu$.  This term in graphene is discussed in Ref.~\onlinecite{Exchange}.
At typical experimental densities, the Fermi temperature 
$T_F \equiv |E_F|/k_B$ in graphene is very high compared to the temperature of the sample ({\em e.g.}, for $n = 10^{13}\,{\rm cm}^{-2}$, $T_F \approx
3.1 \times 10^3\,{\rm K}$, and $T_F\propto \sqrt{n}$).  
%( $T_F = v_0 k_F\hbar/k_B = v_0 \sqrt{\pi n}\hbar/k_B = 10^6 \times
%\sqrt{3.14 \times 10^{16}} \times 1.054 \times 10^{-34}/1.38\times 10^{-23}$\,K) 
Therefore, it is an excellent approximation to set the temperature $T=0$, which we do in the rest of this paper.
This implies that $n_F(\xi_{{\bf k}s}) = \theta(\xi_{{\bf k}s})$ (where $\theta$ is Heaviside unit step function), and $\mu = E_F$. 

The correlation part, $\Sigma^{\rm
cor}_{s}({\bf
  k},\omega)$, is defined to be
the part of $\Sigma_s^{\rm ret}({\bf k},\omega)$ not included in $\Sigma^{\rm ex}_{s}({\bf k})$.
In the $G_0W$ approximation, the $\Sigma^{\rm cor}_{s}({\bf k},\omega)$
can be decomposed into  the line and pole contributions, $\Sigma^{\rm cor} = \Sigma^{\rm line} +
\Sigma^{\rm pole}$,\cite{Quinn58} where
\begin{align}
\Sigma^{\rm line}_s({\bf k},\omega) =
 - \sum_{s'{\bf q}}\int^{\infty}_{-\infty}
\frac{d\omega'}{2\pi} 
&\frac{V_c({\bf q})F_{ss'}({\bf k,k+q})}{\xi_{{\bf
      k+q},s'}-\omega-i\omega'} \nonumber \\
&\times \left [\frac{1}{\epsilon(q,i\omega')} -1 \right]
\end{align}
\begin{align}
\Sigma^{\rm pole}_s&({\bf k},\omega) = \sum_{s'{\bf q}}\left [
  \theta(\omega - \xi_{{\bf k+q},s'})-\theta(-\xi_{{\bf k+q},s'}) \right ]
\nonumber \\
&\times  V_c({\bf q})F_{ss'}({\bf k,k+q})
\left [ \frac{1}{\epsilon(q,\xi_{{\bf k+q},s'}-\omega)}-1 \right ].
\end{align}
The $\Sigma^{\rm ex}$ and $\Sigma^{\rm {line}}$ are completely real, the latter because
$\epsilon(q,i\omega)$ is real. Thus, Im[$\Sigma_s^{{\rm pole}}({\bf k},\omega)$] gives
the total contribution to the imaginary part of the self-energy, {\em i.e.},
\begin{align}
{{\rm Im}}\![\Sigma_s^{\rm ret} &(\bm k,\omega)]\!=\!
\sum_{s'}\!\int\!\frac{d\bm q}{(2\pi)^2} \left [ 
  \theta(\omega-\xi_{\bm k+\bm q,s'}) - \theta(-\xi_{\bm k+\bm q,s'})
\right ] \nonumber \\ 
&\times\, V_c(q)\,{\rm Im}\!\left[\frac{1}{\varepsilon(q,\xi_{\bm k+\bm
      q,s'}-\omega)}\right] F_{ss'}({\bm 
  k},{\bm k}+{\bm q}).
\label{eq:5} 
\end{align}

The inverse quasiparticle lifetime (or, equivalently, the scattering
rate) $\Gamma_s(\bm{k})$ of state $|s\bm k\rangle$  
is obtained by setting the frequency in imaginary part of the
self-energy to the on-shell (bare quasiparticle) energy $\xi_{s\bm
  k}$, {\em i.e.},  
\begin{equation}
\Gamma_s(\bm{k}) = 2\,{{\rm Im}}[\Sigma_s^{\rm ret}(\bm{k},\xi_{\bm
  ks})]. 
\end{equation}
(The factor 2 comes from the squaring of the wavefunction to obtain
the occupation probability.)  The $G_0W$ self-energy approximation
used  here is equivalent to the Born approximation for the scattering
rate.  
Note that the integrand of Eq.~(\ref{eq:5}) is non-zero only when
${\rm Im}[\epsilon]\ne 0$ or ${\rm Re}[\epsilon] = 0$.  These
correspond to scattering off  
single-particle excitations and plasmons, respectively.

%%%%%%%%%%%%%%%%%%%%%%%%---FIGURE 1---%%%%%%%%%%%%%%%%%%%%%%%%%%%%%%
\begin{figure}
\centering
%\subfigure[Parabolic Band]
\includegraphics[scale=0.45]{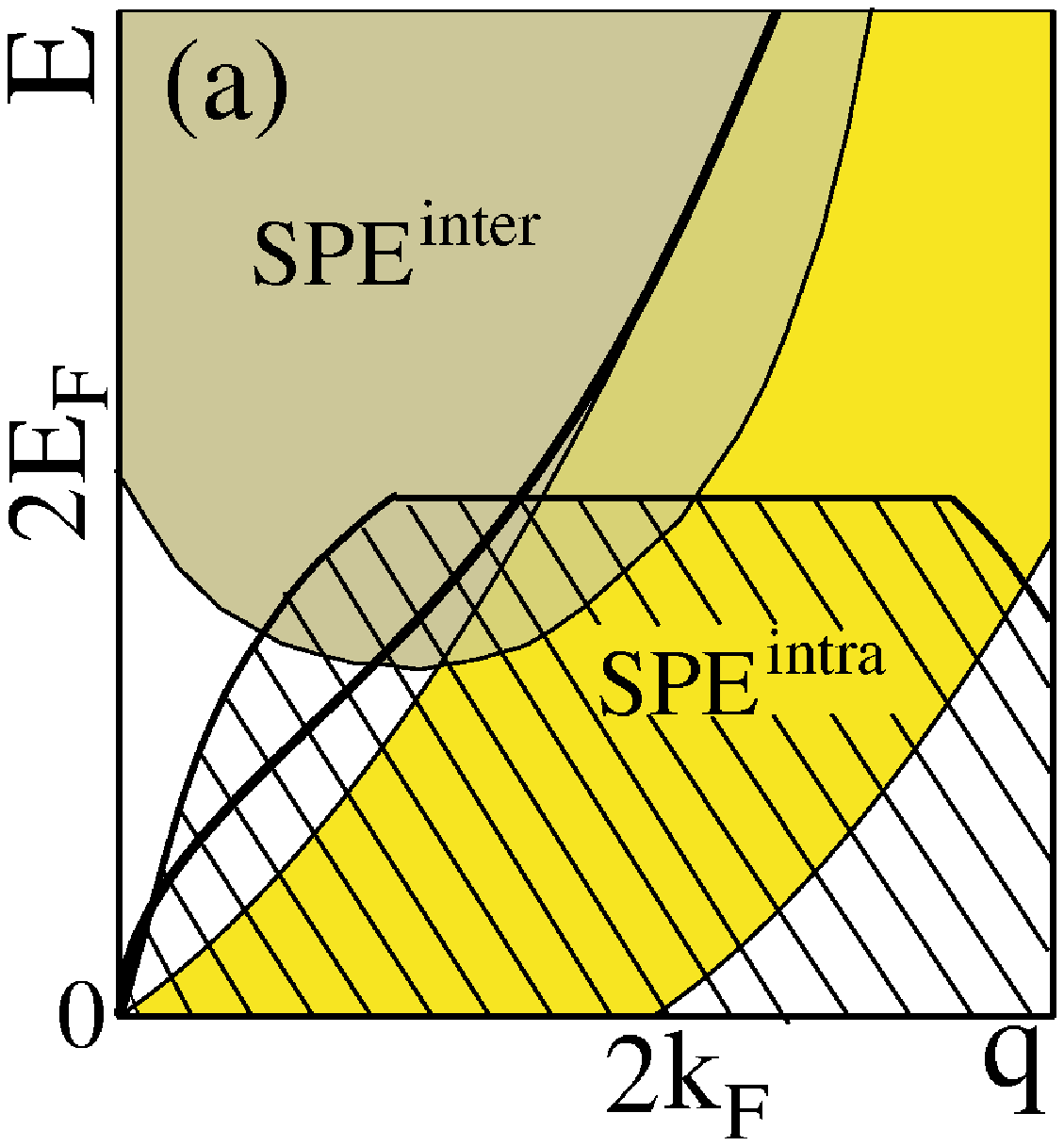}
%\hspace{0.1cm}
%\subfigure[Graphene]
\includegraphics[scale=0.4]{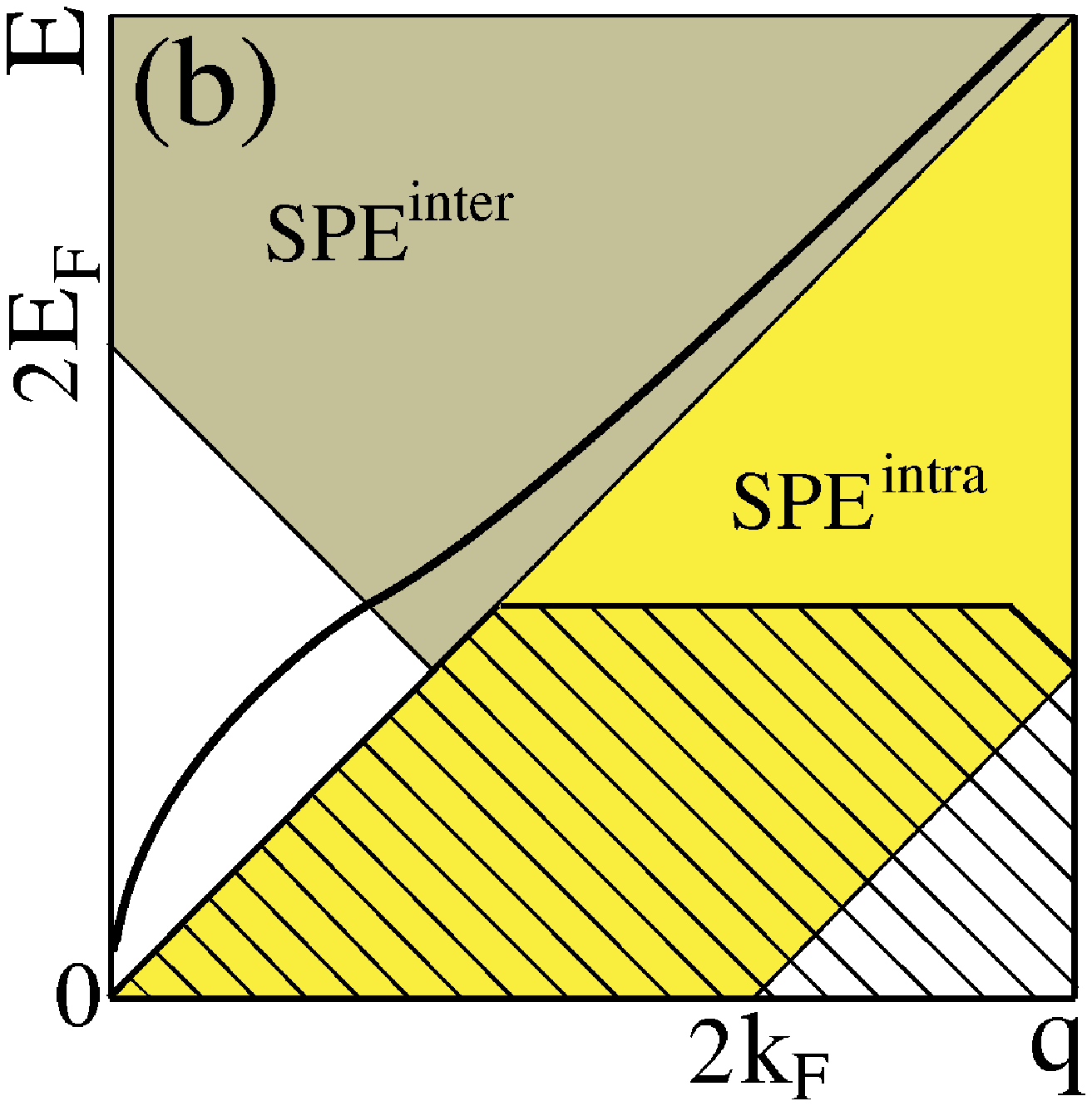}
\caption{(Color online) The single-particle excitations (intraband and
  interband)  
and injected-particle energy-loss (hatched region) continua, and
plasmon dispersion (thick line) for  
(a) gapless parabolic-band semiconductor with equal hole and electron
masses and (b) graphene, 
at $T=0$ with Fermi energy in the the conduction band.  For gapped
semiconductors, the interband continua are shifted up by the energy
gap. 
}
\label{fig:1}
\end{figure}
%%%%%%%%%%%%%%%%%%%%%%%%%%%%%%%%%%%%%%%%%%%%%%%%%%%%%%%%%%%%%%%%%%

\section{quasi-particle  scattering   rate} 

The self-energies and quasiparticle lifetimes of graphene and
conventional parabolic-band semiconductors differ considerably. 
These differences can be explained with the help of Fig.~1, which
shows the single-particle excitation (SPE) and injected-electron
energy loss (IEEL) continua and the plasmon dispersion  
for a direct gapless 2D parabolic band semiconductor and for graphene.
The intersections of the IEEL continua with the SPE continua and the
plasmon  
dispersion lines indicate allowed decay processes via $e$--$e$
interactions. 
In both doped parabolic-band semiconductors and graphene, an injected
electron will decay via single-particle intraband excitations of
electrons in  
the conduction band.  In parabolic band semiconductors, an electron
injected with sufficient kinetic energy can also  
decay via plasmon emissions and interband SPE (also known as ``impact
ionization").  On the other hand, as shown in Fig.~1(b),  
electrons injected into doped graphene 
cannot decay via plasmon emission, and the region in $q$--$\omega$
where decay via interband SPE is allowed 
(along the straight line segment between $\omega = v_0 k_F$ and $v_0
(k-k_F)$, for $k>2k_F$) is of measure zero.   
In fact, within the Born approximation the quasiparticle lifetime of
graphene due to  
$e$--$e$ interactions at $T=0$ is infinite. 
(Multiparticle excitations, which are excluded in the approximations
used here, will give a quasiparticle a finite lifetime
\cite{Guinea_PRL}, but this 
is a relatively small effect in graphene.)

%%%%%%%%%%%%%%%%% Fig 2 %%%%%%%%%%%%%%%%%%
\begin{figure}
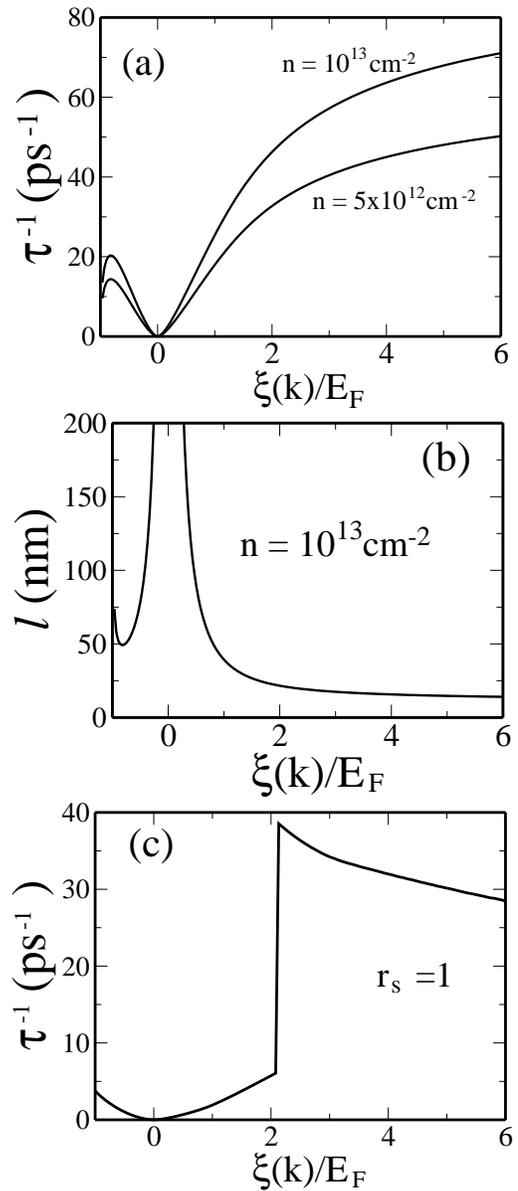

\includegraphics[scale=0.38]{prb_fig2a.eps}
\includegraphics[scale=0.4]{prb_fig2b.eps}
\includegraphics[scale=0.38]{prb_fig2c.eps}
\caption{
(a) Inelastic quasiparticle lifetime/scattering rate
($1/\tau=\Gamma$) in graphene due to dynamically screened $e$--$e$
interactions,  
as a function of energy at $T=0$ for different densities,  within the
Born approximation. (b) The corresponding quasiparticle mean
free path for $n=10^{13}$ cm$^{-2}$ (corresponding to $E_F\approx 0.4\,$eV).
(c) Equivalent scattering rate for a parabolic band semiconductor
(without interband processes), for comparison. 
}
\end{figure}
%%%%%%%%%%%%%%%%%%%%%%%%%%%%%%%%%%%%%%%%%%%%%%%%%

In doped graphene, the only independent parameters relevant for Born
appxoimation quasiparticle scattering rates at $T=0$ are the  
Fermi energy relative to the Dirac point $E_F = v_0 k_F$ and the
dimensionless 
coupling constant $r_s = e^2/(\kappa v_0)$.
%$(4.8\times 10^{-10})^2/(\epsilon \times 10^8
%\times 1.054 \times 10^{-27})
%\approx$  
The scattering rate, which has units of energy, must therefore be
proportional to $E_F$, and must be a function only of $\epsilon/E_F =
k/k_F$ and $r_s$.   
Fig.~2(a) shows the Born approximation $T=0$ quasiparticle
lifetime $1/\tau=\Gamma$ due to $e$--$e$ interactions 
as a function of energy $\xi(k) = \epsilon_k-E_F$. Since the speed of
the quasiparticles close the the Dirac points is approximately a
constant $v_0 \approx 10^8\,{\rm cm/s}$, 
the inelastic mean free path $\ell$ is obtained by
$\ell(\xi) = v_0 \tau(\xi)$.  In Fig.~2(b), we provide the
corresponding $\ell$, which shows that at $n=10^{13}\,{\rm cm}^{-2}$ a
hot electron injected with an energy of 1\,eV above  
$E_F$ has an $\ell$ due to $e$--$e$ interactions that is on
the order of 20\,nm. This will have implications for designing any hot
electron transistor type graphene devices.  In particular, because
Klein 
tunneling\cite{Kleintunnel} in graphene creates problems in the
standard gate-potential switching method in transistors, in its place
could be  
a switch based on modulating the electron energy $\xi$ a regime where
$|\partial\ell/\partial \xi|$ is large.   
 
As with doped parabolic-band 2D semiconductors\cite{ChapHodg,
  Guiliani,DasSarmaMacDonald}, in graphene $\Gamma(k)=1/\tau(k) \propto
(k-k_F)^2\,|\log(|k-k_F|)|$ for $k \approx k_F$  
due to scattering phase-space restrictions.   Further away from $k_F$,
however, the dependence of $\Gamma$ on $k$ in graphene and in
parabolic-band semiconductors are markedly and qualitatively
different. 
To wit, in parabolic band semiconductors plasmon emission
\cite{Guiliani,Jalabert} and interband collision thresholds
\cite{Keldysh} cause discontinuities  
in the $\Gamma(k)$, as shown in Fig.~2(c),  
but the graphene $\Gamma(k)$ is a smooth function because both plasmon
emission and interband processes are absent.

\section{self energy}

In order to see the effects of the plasmons and interband SPE in
graphene in ${\rm Im}[\Sigma^{\rm ret}]$,  
one must look {\em off}-shell, {\em i.e.}, $\omega \ne \xi_{\bm k,s}$. 
The off-shell ${\rm Im}[\Sigma^{\rm ret}_s]$ is not merely of academic
interest; it is  
needed to interpret data from ARPES.   The spectra of the ARPES
electrons ejected  
from graphene give the electronic spectral function, from which one
can infer $\Sigma^{\rm ret}_s(\bm k,\omega)$ \cite{Damascelli}. 
Physically, in the $G_0W$ approximation, the off-shell $2{\rm
  Im}[\Sigma^{\rm ret}_s({\bf k},\omega)]$ gives the  
Born approximation decay rate of the quasiparticle in state $\bm k$ if
it had kinetic energy $\omega+E_F$ rather than $\xi_{\bm k}+E_F$.  

In Fig.~3 we show ${\rm Im}[\Sigma^{\rm ret}_+(k,\omega)]$ for $k=0$ and $k=k_F$ as a function of $\omega$. 
Within the $G_0W$ approximation, the contributions to the off-shell ${\rm Im}[\Sigma^{\rm ret}_s(\bm k,\omega)]$ can be visualized as 
the intersection of the SPE continuum and plasmon line in
Fig.~\ref{fig:1} with  
the vertically displaced IEEL \cite{Offshellsigma}.
At $k=0$ there are two contributions to ${\rm Im}[\Sigma]$, the intraband and interband SPEs.
For low energies ($|\omega| \alt E_F$) only intraband SPE contributes to ${\rm Im}[\Sigma]$. 
Its contribution reaches a maximum around Fermi energy, then decreases
gradually with increasing energy, as is the case with a parabolic-band semiconductor \cite{Jalabert}
where it is the only decay channel for the quasiparticle (assuming $\omega < $ band gap energy). 
But in graphene there is a new decay channel of the quasiparticle,
the interband SPE.  Due to the phase space restrictions the interband SPE
does not contribute to the self energy at low energies, but at higher
energies ($\omega \agt E_F$) its contribution increases sharply, overwhelming the SPE$^\text{intra}$ contribution. 
The SPE$^\text{inter}$ contribution then increases almost linearly with $\omega$, with the same slope 
as for intrinsic graphene \cite{sarma:121406}. %{\bf\large\{not surprisingly since the mathematical form is the same?? - Ben\}}. 
Plasmons do not contribute to ${\rm Im}[\Sigma^{\rm ret}_+(\bm k=0,\omega)]$ for $\omega > 0$. 

%%%%%%%%%%%%%%%%%%%% Fig. 3%%%%%%%%%%%%%%%%%%%%

\begin{figure}
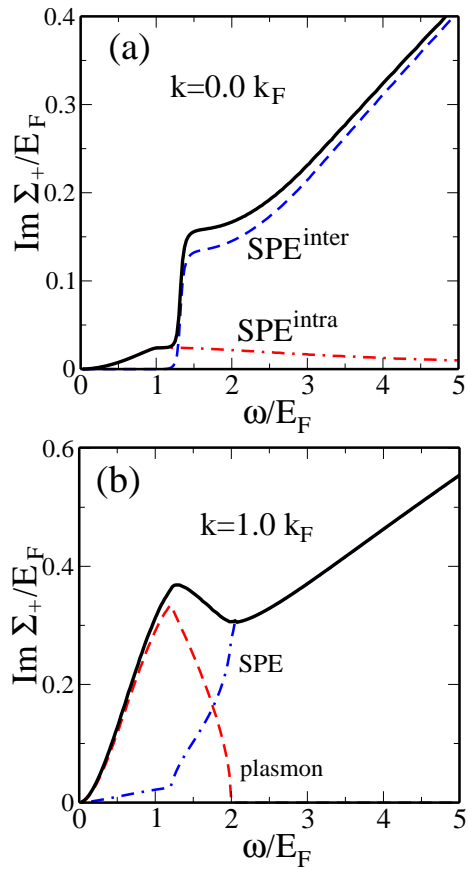

\includegraphics[scale=0.35]{fig3a.eps}
\includegraphics[scale=0.35]{fig3b.eps}
\caption{
(Color online) The imaginary part the retarded self-energy at $T=0$
(thick line) for (a) $k=0$ and (b) 
$k=k_F$ as a function of energy.   The dot-dashed and dashed lines in
(a) are the SPE$^\text{intra}$ and  
SPE$^\text{inter}$ contributions, respectively.  The dot-dashed and the
dashed lines in (b) are the total SPE (intraband and interband SPE) and
the plasmon contributions, respectively.  
}
\end{figure}
%%%%%%%%%%%%%%%%%%%%%%%%%%%%%%%%%%%%%%%%%%%%%

At $k=k_F$, not only do plasmons contribute to ${\rm Im}[\Sigma]$, in the low-energy ($\omega \alt 2E_F$) regime, 
they actually dominate over the SPE contributions, as can be seen in Fig.~3(b). 
In contrast, in parabolic-band 2DEGs the plasmon and SPE contributions to ${\rm Im}[\Sigma^\text{ret}(k_F,\omega)]$ 
go as $\omega^2$ and $\omega^2\ln\omega$, respectively \cite{Guiliani}, and hence both contributions are roughly equal in magnitude.

%%%%%%%%%%%%%%%%% Fig 4 %%%%%%%%%%%%%%%%%%
\begin{figure}
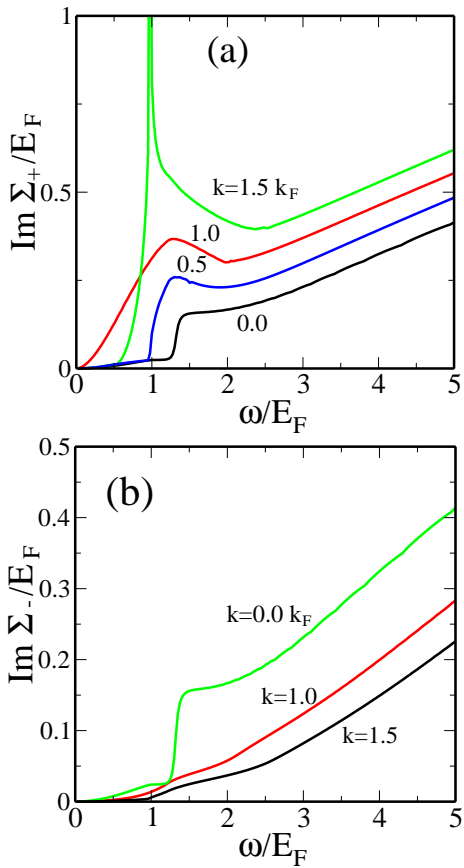

\includegraphics[scale=0.35]{fig4a.eps}
\includegraphics[scale=0.35]{fig4b.eps}
\caption{
(Color online) The imaginary part of the self energy as a
function of energy for wavevectors $k=0$, 0.5, 1.0, 1.5$k_F$ for (a)
the conduction  
band and (b) the valence band of electron-doped graphene at $T=0$. 
}
\end{figure}
%%%%%%%%%%%%%%%%%%%%%%%%%%%%%%%%%%%%%%%%%%%%%

Fig.~4 shows the imaginary part of the quasiparticle self-energies 
for the conduction band, ${\rm Im}[\Sigma_+^\text{ret}]$, and the valence band,
${\rm Im}[\Sigma_-^\text{ret}]$, of graphene for several different wavevectors. For $k>k_F$, the ${\rm Im}[\Sigma_+^\text{ret}]$ 
shows a sharp peak associated with the plasmon emission threshold.
In the high energy regime, the dominant contribution to ${\rm Im}[\Sigma]$ comes from the interband
SPE, which gives rise to an ${\rm Im}[\Sigma]$ is linear in $\omega$ for all wavevectors. 
Note that within the $G_0W$ approximation, for a given $r_s$ in graphene, the $\Sigma$, $\omega$ and $k$ scale with $E_F$, $E_F$ and $k_F$, respectively; 
{\em i.e.}, for fixed $r_s$, the function $\tilde\Sigma(\tilde{k},\tilde{\omega})$ is universal, where $\tilde\Sigma = \Sigma/E_F$,
$\tilde{k}  = k/k_F$ and $\tilde\omega = \omega/E_F$.

%%%%%%%%%%%%%%%%%%%%%%%%%%%%%%%%%%%%%%%%%%%%%%%%%%%%%%%%%%
%%%%%%%%%%%%% Fig. 5
\begin{figure}
\includegraphics[scale=0.4]{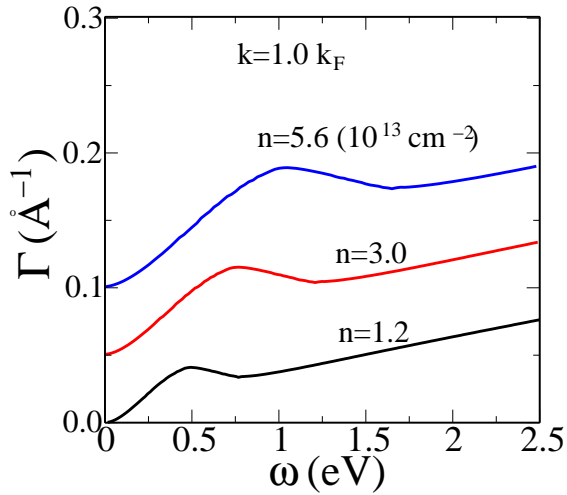}
\caption{
(Color online) Scattering rate as a function of energy for different densities,
$n=1.2$, 3.0, 5.6$\times 10^{13}$ cm$^{-2}$ (for clarity, successive
lines are shifted upward by 0.05 \AA$^{-1}$), in the form of Fig. 3  of 
Bostwick {\it et al.}~\cite{Bostwick}.
}
\end{figure}
%%%%%%%%%%%%%%%%%%%%%%%%%%%%%%%%%%%%%%%%%%%%%%%%%

Finally, Fig.~5 shows the calculated ${\rm Im}[\Sigma_+^\text{ret}]$ at
fixed $k=k_F$ as a function of energy $\omega$ for various densities.
{\it No fitting parameters were used} in this calculation.
These results compare favorably with the recent data from ARPES experiments by Bostwick, {\em et al.}~\cite{Bostwick}
from which they extract ${\rm Im}[\Sigma^{\rm ret}(k_F,\omega)]$ for different densities.  (The overall scale is different because we have
assumed a SiO$_2$ substrate used by some other groups, whereas Ref.~\onlinecite{Bostwick}'s samples were on SiC, which
has a different $\kappa$.) 
Bostwick, {\em et al.}~invoked plasmons, SPE and phonon effects in interpreting their data.  
%Bostwick {\em et al.}\ make the assumption that the $\Sigma(k,\omega)$ is approximately independent of wavevector for $k$ around $k_F$. 
%We have verified that this is indeed true for $0.8 k_F \alt k \alt 1.2 k_F$.
%Note that the energy dependence of ${\rm Im}[\Sigma(k_F,\omega)]$ comes entirely from the SPE and plasmon contributions 
%[see Fig.~3(b)] due to the $e$--$e$ interaction.  
We find that including just the plasmon and SPE effects, we get reasonable agreement with their data, except for features near 0.2\,eV
which probably can be explained by calculations that include electron--phonon interactions.

Before concluding, we discuss some of the approximations that lead to
Eq.~(\ref{eq:1}).  We ignore on-site (Hubbard) $e$--$e$ interactions 
because, at zero magnetic field, this interaction is irrelevant in the
renormalization group sense\cite{PhysRevB.63.134421}.  
Our calculation {\em does} include both intra- and inter-band scattering processes,
since our expression for $\Sigma_s({\bf k},i\omega_n)$, Eq.~(\ref{eq:1}), 
contains a pseudospin sum $s' = \pm 1$. 
Finally, there is an issue about the coupling of the bands due to interactions;  {\em i.e.}, because of 
non-diagonal terms $\Sigma_{ss'}$ with $s\ne s'$, the elementary excitations are a
superposition of states from the conduction and the  
valence band.  Dyson's equation in this case is 
${\bf G}^{-1} = {\bf G}_0^{-1} - {\bf \Sigma}$, where the quantities
are $2\times 2$ matrices, which, when written out in full, is 
\begin{align}
&{\bf G}^{-1}({\bf k},\omega) = \nonumber\\
&\left( \begin{array}{cc}
\omega - \xi_+(k) - \Sigma_{++}({\bf k},\omega) & -\Sigma_{+-}({\bf k},\omega)\\
-\Sigma_{-+}({\bf k},\omega) & \omega - \xi_{-}(k) - \Sigma_{--}({\bf k},\omega)
\end{array} \right).
\end{align}
The excitation energies are given by the poles of ${\bf G}$, or equivalently, the zeroes of the determinant of ${\bf G}^{-1}$. 
Our calculation is equivalent to ignoring the off-diagonal terms $\Sigma_{+-}\Sigma_{-+}$ in the determinant of ${\bf G}^{-1}$.
%$\omega - \epsilon_+(k) - \Sigma_{++} = 0$ and $\omega -
%\epsilon_{-}(k) - \Sigma_{--} = 0$.  
The off-diagonal terms $\Sigma_{\pm\mp}$ are each of order $r_s$, and therefore 
the product $\Sigma_{+-}\Sigma_{-+}\sim r_s^2$ which can be ignored in the weak-coupling regime.
We also ignore intervalley scattering 
because the Born approximation Coulomb scattering rate in 2D is
$\propto q^{-2}$, implying that small-$q$ intravalley scattering
processes dominate over 
the large-$q$ intervalley ones.

\section{conclusion}

To conclude, we have calculated the electron-electron interaction
induced hot electron inelastic scattering in graphene, finding a
number 
of intriguing and significant differences with the corresponding 2D
parabolic dispersion systems.  Our infinite ring-diagram $G_0W$
appracomation 
should be an excellent quantitative approximation for graphene since
graphene is a low-$r_s$ ({\em i.e.}, weak-coupling) system.  
Our $T=0$ results should remain 
for $T \ll T_F$, which is typically the case in experiments. 
We obtain good agreement with recent ARPES data without invoking any
phonon effects. 
ARPES data has recently created a controversy with respect to the role of
plasmons \cite{Bostwick,Lanzara}.
Our detailed calculation using a realistic model  
generally agrees with the interpretation of the data in
Ref.~\onlinecite{Bostwick}.  
The calculated inelastic scattering length $\ell$ as a
function of energy $\xi$ is by itself is of interest in the context
of ballistic hot electron transistor applications of graphene,  
where the performance limitation is inelastic scattering.
Furthermore, the Klein tunneling effect in graphene makes it difficult  
to switch transistors by a gate-voltage induced depletion (as in a
conventional MOSFET).  In its place, one could imagine a 
graphene-based transistor in which switching is achieved by modulation
of the injection carrier energy in the regime of $\xi$  
where $|d\ell(\xi)/d\xi|$ is large.

This work is supported by US-ONR.

%\bibliography{../self_im}

\begin{thebibliography}{17}
\bibitem[{Off()}]{Offshellsigma}
\bibinfo{note}{More accurately, the contributions come from the regions in the
  $E$--$q$ plane between $E=0$ and $E=\omega$ in which the SPE and plasmon
  lines intersect with the energy loss {\em and} gain spectrum that is shifted
  along the $E$-axis by $\omega -\xi_{\bm k}$.}
\end{thebibliography}
%\end{document}

\end{document}